\documentstyle[aps,pra,epsfig,eqsecnum]{revtex}
\title{Quantum copying: A network }
\author{
V. Bu\v{z}ek$^{1,2}$, S.L. Braunstein$^{3}$,   M. Hillery$^{4}$,
and D. Bru\ss$^{5}$}
\address{
$^{1}$ Optics Section, The Blackett Laboratory,
Imperial College, London SW7 2BZ,  UK\newline
$^{2}$ Institute of Physics, Slovak Academy of Sciences, Dubravsk\'a
cesta 9, 842 28 Bratislava, Slovakia\newline
$^{3}$ SEECS, University of Wales, Dean Street, Bangor LL57 1UT, UK,\newline
$^{4}$Department
of Physics and Astronomy, Hunter College, CUNY,
695 Park Avenue, New York, NY 10021, USA\newline
$^{5}$ Clarendon Laboratory, Department of Physics, University of Oxford,
Parks Road, Oxford OX1 3PU, UK
}
\date{March 19, 1997}

\begin{document}

\maketitle
\begin{abstract}
We present a network consisting of quantum gates which produces two
imperfect copies of an arbitrary qubit. The quality of the copies
does not depend on the input qubit.  We also show that for a restricted
class of inputs it is possible to use a very similar network to
produce three copies instead of two.  For qubits in this class, the copy
quality is again independent of the input and is the same as the
quality of the copies produced by the two-copy network.
\end{abstract}
%%%%%%%%%%%%%%%%%%%%%%%%%%%%%%%%%%%%%%%%%%%%%%%%%%%%%%%%%%%%%%%%%
%\begin{multicols}{2}
%\narrowtext
%%%%%%%%%%%%%%%%%%%%%%%%%%%%%%%%%%%%%%%%%%%%%%%%%%%%%%%%%%%%%%%%%%%%%%%%%%

\section{INTRODUCTION}
Since the work of Wooters and Zurek it has been known that it is
impossible  to copy (i.e., clone)  perfectly
an arbitrary quantum state \cite{Wootters,Barnum}.
  These authors considered a quantum copy
machine which is supposed to copy a qubit and demonstrated that if it
copies
two basis vectors correctly, it cannot copy superpositions of these
vectors
without introducing errors.  This result follows directly from the fact
that
quantum mechanical transformations are implemented by linear operators.

If one is only interested in producing imperfect copies, however, then it
is possible to design machines (actually, find unitary transformations)
which copy quantum states.  A number of these were analyzed in a recent
paper by two of us \cite{Buzek}.
The copy machine considered by Wooters and
Zurek, for example, produces two identical copies at its output, but
the quality of these copies depends upon the input state.  They
are perfect for the basis vectors which we denote as $|0\rangle$
and $|1\rangle$, but, because the copying process
destroys the off-diagonal information of the input density matrix,
they are poor for input states of the form $(|1\rangle +
e^{i\varphi}|0\rangle )/\sqrt{2}$, where $\varphi$ is arbitrary.  A different
copy machine, the Universal Quantum Copy Machine (UQCM), produces two
identical copies whose quality is independent of the input state.  In
addition, its performance is, on average, better than that of the
Wooters-Zurek machine, and the action of the machine simply scales the
expectations values of certain operators.  In particular the expectation
value in one of the copies of any operator which is a linear combination
of the Pauli matrixes is $2/3$ that of its expectation value in the input
state.	Gisin has recently generalized the UQCM for the cases in which
there are $N$ identical inputs and $N+1$ outputs, that is one copy is
produced, and also in which there are $N$ inputs and $N+2$ outputs,
i.\ e.\ there are two copies produced \cite{Gisin}.
In both cases all of the
output copies are identical and their fidelity, that is their overlap
with the input state, goes to $1$ as $N$ goes to infinity.

In this paper we want to do two things.  First, we present  a
quantum logic network which realizes the UQCM.	An analysis of this
network suggests that it should be possible to produce not two
(imperfect)
copies of the input state at the output, but three.  Second, we find that
a very similar quantum network, can also be used as a
quantum ``triplicator'', i.e. a copying machine which produces three
(imperfect) copies of the original qubit. In general,
the triplicator has the undesirable feature that
the quality of the copies which emerge from it is  state-dependent.
However, if the original qubit is in a
superposition state $\alpha |0\rangle+\beta|1\rangle$ with $\alpha$ and
$\beta$
{\it real} then the quality of the copied qubits
does not depend on the particular value of $\alpha$. Moreover we show
that in this case the quality of the triplicated qubits
is the same as those which emerge
from the UQCM, which is a ``duplicator''.

In addition, we discuss the quantum
entanglement of the qubits at the output of our quantum copying networks.
The fact that the copies are entangled means that they are
not independent; measuring
one copy can have an effect on the other.  This feature is
something which must be kept in mind when determining how
to make use of the copies.

The quantum logic networks which we propose
consist of one and two-bit quantum gates for
which proposed designs already
exist.	They should, therefore, be useful in the experimental
realization of quantum copy machines.

The paper is organized as follows. In Section II we briefly review
the unitary transformation which specifies the UQCM. The
quantum copying networks are
described in Section III, while in Section IV we discuss the inseparability
of the copied qubits. The quantum triplicator is described in Section V.

\section{UNIVERSAL QUANTUM COPY MACHINE}
Let us assume we want to copy
an arbitrary pure state $|\Psi\rangle_{a_1}$ which in a particular
 basis
$\{|0\rangle_{a_1},|1\rangle_{a_1}\}$ is described by
the state vector $|\Psi\rangle_{a_1}$
\begin{eqnarray}
|\Psi\rangle_{a_1} = \alpha |0\rangle_{a_1} +\beta |1\rangle_{a_1};
\qquad \alpha=\sin\vartheta {\rm e}^{i\varphi};~~~\beta=\cos\vartheta.
\label{1.1}
\end{eqnarray}
The two numbers which characterize the state (\ref{1.1}) can be
associated with the ``amplitude'' $|\alpha|$
and the ``phase'' $\varphi$ of the qubit. Even though
 ideal copying, i.e., the transformation
\begin{eqnarray}
|\Psi\rangle_{a_1} \longrightarrow |\Psi\rangle_{a_1} |\Psi\rangle_{a_2}
\label{1.2}
\end{eqnarray}
is prohibited by the laws  of quantum mechanics for an {\em arbitrary}
state (\ref{1.1}), it is still possible to design quantum copiers which
operate reasonably well. In particular, the UQCM \cite{Buzek}
is specified by the following conditions.

{\bf (i)}The  state of the original system and its quantum copy at the
output of the quantum copier, described by density operators
$\hat{\rho}^{(out)}_{a_1}$ and $\hat{\rho}^{(out)}_{a_2}$, respectively,
are identical, i.e.,
\begin{eqnarray}
\hat{\rho}^{(out)}_{a_1}   = \hat{\rho}^{(out)}_{a_2}
\label{1.3}
\end{eqnarray}

{\bf (ii)} If no {\em a priori} information about the {\em in}-state
of the original system is available, then it is reasonable to require
that {\em all} pure states should be copied equally well. One way to
implement
this assumption is to design a quantum copier such that the distances between
density operators of each system at the output ($\hat{\rho}^{(out)}_{a_j}$
where $j=1,2$)	and the ideal density operator $\hat{\rho}^{(id)}$
which describes the {\em in}-state of the original mode are input state
independent.  Quantitatively this means that if we employ the square of
the Hilbert-Schmidt norm
\begin{eqnarray}
d(\hat{\rho}_{1};\hat{\rho}_{2}):= {\rm Tr}\left[\left(
\hat{\rho}_{1}-\hat{\rho}_{2}\right)^2\right],
\label{1.4}
\end{eqnarray}
as a measure of distance between two operators, then the quantum copier
should be such that
\begin{eqnarray}
d_{1}(\hat{\rho}_{a_j}^{(out)};\hat{\rho}_{a_j}^{(id)})=
{\rm const.};\qquad j=1,2.
\label{1.5}
\end{eqnarray}
Here we use the subscript 1 in the definition of the distance $d_1$, to
signify that this is the distance between  single-qubit states.

{\bf (iii)}
Finally, we would also like to require that
the copies are as close as possible to the ideal output state,
which is, of course, just the input state.
This means that we want our quantum copying transformation
to satisfy
\begin{eqnarray}
d_1(\hat{\rho}_{a_j}^{(out)};\hat{\rho}_{a_j}^{(id)}) =
{\rm min}
\left\{d_1(\hat{\rho}_{a_j}^{(out)};\hat{\rho}_{a_j}^{(id)})\right\};
 \qquad (j=1,2).
\label{1.6}
\end{eqnarray}
Originally, the UQCM was found by guessing a transformation which
contained two free parameters, and then determining them by
demanding that condition (ii) be satisfied, and that the
distance between the two-qubit output density matrix and the
ideal two-qubit output be input state independent.  That the
UQCM machine obeys the condition (\ref{1.6}) has only been shown
recently by one of us \cite{Bruss} .

The unitary transformation which implements the
UQCM \cite{Buzek} is given by
\begin{eqnarray}
|0\rangle_{a_1} |Q\rangle_{x} &\rightarrow & \sqrt{\frac{2}{3}}|00
\rangle_{a_1a_2}|\uparrow\rangle_{x}+\sqrt{\frac{1}{3}}|+\rangle_{a_1a_2}
|\downarrow\rangle_{x} \nonumber \\
|1\rangle_{a_1} |Q\rangle_{x} &\rightarrow & \sqrt{\frac{2}{3}}|11
\rangle_{a_1a_2}|\downarrow\rangle_{x}+\sqrt{\frac{1}{3}}|+\rangle_{a_1a_2}
|\uparrow\rangle_{x},
\label{1.7}
\end{eqnarray}
where
\begin{equation}
|+\rangle_{a_1a_2} = \frac{1}{\sqrt{2}}(|10\rangle_{a_1a_2}+
|01\rangle_{a_1a_2}),
\label{1.8}
\end{equation}
and satisfies the conditions (\ref{1.3}-\ref{1.6}).
The system labelled by $a_1$ is the original (input) qubit, while
the other system $a_2$ represents the qubit onto which the information
is copied. This qubit is supposed to be initially in a state
$|0\rangle_{a_2}$ (``blank paper'' in a copier).
The states of the copy machine
are labelled by $x$.  The state space of the copy machine is two
dimensional, and we assume that it is always in the same state
$|Q\rangle_{x}$ initially.  If the original qubit is in the
superposition state  (\ref{1.1})
then the reduced density operator of both copies
at the output are equal [see condition (\ref{1.3})] and they can be
expressed as
\begin{equation}
\hat{\rho}_{a_j}^{(out)}
=\frac{5}{6}|\Psi\rangle_{a_j}\langle\Psi|+
\frac{1}{6}|\Psi_{\perp}\rangle_{a_j}\langle\Psi_{\perp}|,\qquad j=1,2
\label{1.9}
\end{equation}
where
\begin{equation}
|\Psi_{\perp}\rangle_{a_j}=\beta^{\star} |0\rangle_{a_j}-
\alpha^{\star} |1\rangle_{a_j} ,
\label{1.10}
\end{equation}
is the state orthogonal to $|\Psi\rangle_{a_j}$.  This
implies that the copy contains $5/6$ of the state we want and $1/6$
of that one we did not.

We note that the density operator $\rho_{a_j}^{(out)}$ given by
Eq.(\ref{1.9})
can be rewritten in a ``scaled'' form:
\begin{equation}
\hat{\rho}_{a_j}^{(out)}
= s_j \hat{\rho}_{a_j}^{(id)}  + \frac{1-s_j}{2} \hat{1};\qquad j=1,2,
\label{1.11}
\end{equation}
which guarantees that the distance (\ref{1.4}) is input-state independent,
i.e. the condition (\ref{1.5} is automatically fulfilled. The scaling
factor in Eq.(\ref{1.11}) is $s_j=2/3$.

\section{COPYING NETWORK}

In what follows we show how with simple quantum logic gates we can copy
quantum information encoded in the original qubit onto other qubits.
The copying procedure can be understood as a ``spread'' of information
via a ``controlled'' entanglement between the original qubit and the
copy qubits. This controlled entanglement is implemented by a sequence
of controlled-NOT operations operating on the original qubit and the
copy qubits which are initially prepared in a specific state.

In designing a network for the UQCM we first note that since
the state space of the copy machine itself is two dimensional, we can
consider it to be an additional qubit.	Our network, then, will take
3 input qubits, one for the input, one which becomes one of the copies,
and one for the machine, and transform them into 3 output qubits two
of which will be copies of the output. In what follows we will denote
the quantum copier qubit as $a_3$ rather than $x$.

The operation of this network will be slightly different from what was
indicated in the previous paragraph.  Rather than have the copies
appear in the $a_1$ and the $a_{2}$ qubit, they will appear in the
$a_{2}$ and $a_{3}$ qubits.

Before proceeding with the network itself let us specify the one and
two-qubit gates from which it will be constructed.
Firstly we define a single-qubit rotation $\hat{R}_j(\theta)$
($j=1,2,3$) which acts on the basis vectors of qubits as
\begin{eqnarray}
\begin{array}{rcl}
\hat{R}_{j}(\theta) | 0 \rangle_j  & = &
\cos\theta | 0 \rangle_j + \sin\theta | 1 \rangle_j;\\
\hat{R}_{j}(\theta) | 1 \rangle_j  & = &
-\sin\theta | 0 \rangle_j + \cos\theta | 1 \rangle_j.
\end{array}
\label{2.1}
\end{eqnarray}

We also will utilize a two-qubit operator (a two-bit quantum gate),
the so-called
controlled-NOT	which has as its inputs a control qubit
(denoted as $\bullet$ in Fig.1) and
a target qubit	(denoted as $\circ$ in Fig.1).
The control qubit
is unaffected by the action of the
gate, and if the control qubit is $|0\rangle$, the target qubit is
unaffected as well.  However, if the control qubit is in the
$|1\rangle$ state, then a NOT operation is performed on the target
qubit.	The operator which implements this gate, $\hat{P}_{kl}$,
acts on the basis vectors of the two qubits as follows ($k$
denotes the control qubit and $l$ the target):
\begin{eqnarray}
\begin{array}{rcl}
\hat{P}_{kl} | 0 \rangle_k | 0 \rangle_l & = &
| 0 \rangle_k | 0 \rangle_l;\\
\hat{P}_{kl} | 0 \rangle_k | 1 \rangle_l & = &
| 0 \rangle_k | 1 \rangle_l;\\
\hat{P}_{kl} | 1 \rangle_k | 0 \rangle_l & = &
| 1 \rangle_k | 1 \rangle_l;\\
\hat{P}_{kl} | 1 \rangle_k | 1 \rangle_l & = &
| 1 \rangle_k | 0 \rangle_l.
\end{array}
\label{2.2}
\end{eqnarray}

We can decompose the quantum copier network into two parts.
In the first part the replica
qubits $a_2$ and $a_3$	are prepared in a specific state
$|\Psi\rangle_{a_2a_3}^{(prep)}$. Then in the second part
of the copying network the original information from the
original qubit
is {\em redistributed} among the three qubits.	  That
is the action of the quantum copier can be described as
a sequence of two unitary transformations
\begin{eqnarray}
|\Psi\rangle_{a_1}^{(in)} |0\rangle_{a_2} |0\rangle_{a_3}
\longrightarrow  |\Psi\rangle_{a_1}^{(in)} |\Psi\rangle_{a_2 a_3}^{(prep)}
\longrightarrow  |\Psi\rangle_{a_1a_2 a_3}^{(out)}.
\label{2.3}
\end{eqnarray}
The network for the quantum copying machine is displayed
in Fig.\ 1.

\subsection{Preparation of quantum copier}
Let us first look at the preparation stage.
Prior to any interaction with the input qubit we have to
prepare the two quantum copier qubits ($a_2$ and $a_3$)
in a very specific state $|\Psi\rangle_{a_2a_3}^{(prep)}$.
If we assume that initially these two qubits are
in the state
\begin{eqnarray}
|\Psi\rangle_{a_2a_3}^{(in)} = |0\rangle_{a_2}	|0\rangle_{a_3},
\label{2.4}
\end{eqnarray}
then the arbitrary state  $|\Psi\rangle_{a_1a_2}^{(prep)}$
\begin{eqnarray}
|\Psi\rangle_{a_2a_3}^{(prep)}= C_1|00\rangle_{a_2a_3}
+C_2|01\rangle_{a_2a_3}+C_3|10\rangle_{a_2a_3}+C_4|11\rangle_{a_2a_3},
\label{2.5}
\end{eqnarray}
with real amplitudes $C_i$ (such that  $\sum_{i=1}^{4} C_i^2=1$)
can be prepared by a simple quantum network (see the ``preparation''
box in Fig.1) with two controlled-NOTs $\hat{P}_{kl}$
and three  rotations $\hat{R}(\theta_j)$, i.e.
\begin{eqnarray}
|\Psi\rangle_{a_2 a_3}^{(prep)}=
\hat{R}_2(\theta_3) \hat{P}_{32}
\hat{R}_3(\theta_2) \hat{P}_{23}
\hat{R}_2(\theta_1)|0\rangle_{a_2}  |0\rangle_{a_3}.
\label{2.6}
\end{eqnarray}
Comparing Eqs.(\ref{2.5}) and (\ref{2.6}) we find a set of equations
\begin{eqnarray}
\begin{array}{rcl}
~\cos\theta_1 \cos\theta_2\cos\theta_3
+\sin\theta_1 \sin\theta_2\sin\theta_3	 & = & C_1;\\
-\cos\theta_1 \sin\theta_2\sin\theta_3
+\sin\theta_1 \cos\theta_2\cos\theta_3	 & = & C_2;\\
~\cos\theta_1 \cos\theta_2\sin\theta_3
-\sin\theta_1 \sin\theta_2\cos\theta_3	 & = & C_3;\\
~\cos\theta_1 \sin\theta_2\cos\theta_3
+\sin\theta_1 \cos\theta_2\sin\theta_3	 & = & C_4,
\end{array}
\label{2.7}
\end{eqnarray}
from which the angles $\theta_j$ (j=1,2,3) of  rotations
can be specified as functions of parameters $C_i$.
In particular, for the purpose of the UQCM we need
that
\begin{eqnarray}
|\Psi\rangle_{a_2a_3}^{(prep)} =\frac{1}{\sqrt{6}}\left(
2 |00\rangle_{a_2a_3} + |01\rangle_{a_2a_3} + |10\rangle_{a_2a_3}\right).
\label{2.8}
\end{eqnarray}
With the help of Eq.(\ref{2.7})
we find that the rotation angles necessary for the preparation
of the state given in Eq.(\ref{2.8}) are
\begin{eqnarray}
\theta_1 = \theta_3 = \frac{\pi}{8};\qquad
\theta_2=-{\rm arcsin}\left(\frac{1}{2}-\frac{\sqrt{2}}{3}\right)^{1/2}.
\label{2.9}
\end{eqnarray}

\subsection{Quantum copying}

Once the qubits of the quantum copier are properly prepared
then the copying of the initial  state $|\Psi\rangle_{a_1}^{(in)}$
of the original qubit can be performed
by a sequence of four controlled-NOT operations (see
Fig. 1)
\begin{eqnarray}
|\Psi\rangle_{a_1a_2a_2}^{(out)} = \hat{P}_{a_3a_1} \hat{P}_{a_2a_1}
\hat{P}_{a_1a_3}
\hat{P}_{a_1a_2}|\Psi\rangle_{a_1}^{(in)}  |\Psi\rangle_{a_2a_3}^{(prep)}.
\label{2.10}
\end{eqnarray}
When this operation is combined with the preparation stage, we find
that the basis states of the original qubit ($a_1$) are
copied as
\begin{eqnarray}
|0\rangle_{a_1}|00\rangle_{a_2a_3} \longrightarrow
\sqrt{\frac{2}{3}}|0\rangle_{a_1} |00\rangle_{a_2a_3} +
\frac{1}{\sqrt{3}}|1\rangle_{a_1} |+\rangle_{a_2a_2};
\label{2.11}
\end{eqnarray}
\begin{eqnarray}
|1\rangle_{a_1}|00\rangle_{a_2a_3} \longrightarrow
\sqrt{\frac{2}{3}}|1\rangle_{a_1} |11\rangle_{a_2a_3} +
\frac{1}{\sqrt{3}}|0\rangle_{a_1} |+\rangle_{a_2a_2},
\label{2.12}
\end{eqnarray}
where
$|+\rangle_{a_2a_3}=(|01\rangle_{a_2a_3}+|10\rangle_{a_2a_3})/\sqrt{2}$.
When the original qubit is in the superposition state (\ref{1.1})
then the state vector of the three qubits after the copying
has been performed reads
\begin{eqnarray}
|\Psi\rangle_{a_1a_2a_3}^{(out)}= |0\rangle_{a_1} |\Phi_0\rangle_{a_2a_3}
+ |1\rangle_{a_1}|\Phi_1\rangle_{a_2a_3},
\label{2.13}
\end{eqnarray}
with
\begin{eqnarray}
|\Phi_0\rangle_{a_2a_2}=\alpha\sqrt{\frac{2}{3}}|00\rangle_{a_2a_3}
+\beta \frac{1}{\sqrt{3}}|+\rangle_{a_2a_3}; \qquad
|\Phi_1\rangle_{a_2a_3}=\beta\sqrt{\frac{2}{3}}|11\rangle_{a_2a_3}
+\alpha \frac{1}{\sqrt{3}}|+\rangle_{a_2a_3}.
\label{2.14}
\end{eqnarray}
From this it follows that at
the output of the  quantum copier we find a pair of entangled
qubits in a state described by the density operator
\begin{eqnarray}
\hat{\rho}_{a_2a_3}^{(out)}= |\Phi_0\rangle_{a_2a_3}\langle \Phi_0|+
|\Phi_1\rangle_{a_2a_3}\langle \Phi_1|.
\label{2.15}
\end{eqnarray}
Each of the copy qubits at the output of the quantum copier
has a reduced density operator $\hat{\rho}_{a_j}^{(out)}$
($j=2,3$)  given by  Eq.(\ref{1.11}). The distance
$d_1(\hat{\rho}_{a_j}^{(out)};\hat{\rho}_{a_j}^{(id)})$ ($j=2,3$)
between the output qubit and the ideal qubit is constant and
can expressed as a function of the scaling parameter $s$
in Eq.(\ref{1.11}):
\begin{eqnarray}
d_1(\hat{\rho}_{a_j}^{(out)};\hat{\rho}_{a_j}^{(id)})
=\frac{(1-s)^2}{2}=\frac{1}{18}.
\label{2.16}
\end{eqnarray}
Analogously we find the distance
$d_2(\hat{\rho}_{a_2a_3}^{(out)};\hat{\rho}_{a_2a_3}^{(id)})$
between the  two-qubit output of the quantum copying and the
ideal output to be constant, i.e.
\begin{eqnarray}
d_2(\hat{\rho}_{a_2a_3}^{(out)};\hat{\rho}_{a_2a_3}^{(id)})
=\frac{s^2}{2}=\frac{2}{9}.
\label{2.17}
\end{eqnarray}

The original qubit after the copying is performed is in
a state
\begin{eqnarray}
\hat{\rho}_{a_1}^{(out)}=
\frac{1}{3}\left(\hat{\rho}_{a_1}^{(in)}\right)^{\rm T} + \frac{1}{3}\hat{1},
\label{2.18}
\end{eqnarray}
where the superscript T denotes the transpose.
We note that in spite of the fact, that
the distance between this density operator and the ideal qubit
depends on the initial state of the original qubit, i.e.
\begin{eqnarray}
d_1(\hat{\rho}_{a_1}^{(out)};\hat{\rho}_{a_1}^{(id)})
=\frac{2}{9}\left(1+ 12 |\alpha|^2|\beta|^2 \sin^2\varphi\right),
\label{2.19}
\end{eqnarray}
the output state of the original qubit still contains
information about the input state, though less than either of the
copies.  In order to extract this information we note that for an
hermitian operator $\hat{A}$
\begin{equation}
{\rm Tr}(\hat{\rho}^{(in)}_{a_1}\hat{A}) =
{\rm Tr}\left( (\hat{\rho}^{(in)}_{a_1})^{\rm T} \hat{A}^{\rm T} \right).
\end{equation}
This means that to obtain information about $\hat{A}$ at the
input, we measure $\hat{A}^{\rm T}$ for the original qubit
at the output.

\section{INSEPARABILITY OF COPIED QUBITS}

An ideal copy machine would produce two copies which
are completely independent of each other, i.\ e.\ the reduced
density matrix for the two copies, $\hat{\rho}_{a_{2}a_{3}}$,
would be a product of $\hat{\rho}_{a_{2}}$ and $\hat{\rho}_{a_{3}}$.
For the UQCM, however, is not the case and there are correlations
between the copies.  These correlations can be either quantum
mechanical or classical, and we would like to determine whether
the two copies are quantum-mechanically entangled.
To do so, we first recall that
a density operator of two subsystems is
inseparable if it {\em cannot} be written as the convex sum
\begin{eqnarray}
	\hat{\rho}_{a_{2}a_{3}}  = \sum_m w^{(m)} \hat{\rho}_{a_{2}}^{(m)}
	\otimes \hat{\rho}_{a_{3}}^{(m)}.
\label{3.1}
\end{eqnarray}
Inseparability is one of the most fundamental quantum phenomenon, which,
in particular, may result in the violation of Bell's inequality (to be
specific, a separable system always satisfy Bell's inequality, but the
contrary is not necessarily true). Note that distant parties
cannot prepare an inseparable state from a separable state if they only
use local operations and classical communication channels \cite{Peres}.

In the case of two qubits (i.e., spins-1/2)
we can utilize the Peres-Horodecki theorem
\cite{Peres,Horodecki-sep}
which states that the positivity of the
partial transposition of a state is both a
{\em necessary} and {\em sufficient} condition for
its separability. Before we proceed further we briefly described how to
``use'' this theorem: The density matrix associated with the density operator
of two spins-1/2 can be written as
\begin{eqnarray}
	{\rho}_{m\mu,n\nu}  = \langle e_m|\langle f_\mu|\hat{\rho}
	| e_n\rangle|f_\nu\rangle,
\label{3.2}
\end{eqnarray}
where $\{ |e_m\rangle\}$ ($\{|f_{\mu}\rangle\}$) denotes an orthonormal
basis in
the Hilbert  space of the first (second) spin-1/2 (for instance,
$|e_0\rangle = |0\rangle_{a_2}$; $|e_1\rangle = |1\rangle_{a_2}$,  and
$|f_0\rangle = |0\rangle_{a_3}$; $|f_1\rangle = |1\rangle_{a_3}$). The
partial transposition  $\hat{\rho}^{T_2}$ of $\hat{\rho}$ is defined as
\begin{eqnarray}
	{\rho}_{m\mu,n\nu}^{T_2}  = {\rho}_{m\nu,n\mu}.
\label{3.3}
\end{eqnarray}
Then the necessary and sufficient condition for the state
$\hat{\rho}_{a_2a_3}$
of two spins-1/2 to be inseparable is that at least one of the eigenvalues
of the partially transposed operator ${\rho}_{m\mu,n\nu}^{T_2}$
is negative.

Now we
 will check whether the density operator
$\hat{\rho}_{a_{2}a_{3}}^{(out)}$ given by Eq.(\ref{2.15}) is separable.
In the basis $\{ |11\rangle_{a_2a_3}, |10\rangle_{a_2a_3},
|01\rangle_{a_2a_3}, |00\rangle_{a_2a_3}\}$ this density
operator is described by a matrix
\begin{eqnarray}
\hat{\rho}_{a_2a_3}^{(out)} =\frac{1}{6}
\left(
\begin{array}{cccc}
4 |\beta|^2 & 2 \alpha^{\star}\beta & 2 \alpha^{\star}\beta & 0\\
2 \alpha\beta^{\star} & 1 & 1 & 2 \alpha^{\star}\beta \\
2 \alpha\beta^{\star} & 1 & 1 & 2 \alpha^{\star}\beta \\
0 & 2 \alpha\beta^{\star} & 2 \alpha\beta^{\star} & 4 |\alpha|^2
\end{array}
\right),
\label{3.4}
\end{eqnarray}
while the corresponding partially transposed operator
in the matrix representation reads
\begin{eqnarray}
\hat{\rho}_{a_2a_3}^{T_2} =
\frac{1}{6}
\left(
\begin{array}{cccc}
4 |\beta|^2 & 2 \alpha\beta^{\star} & 2 \alpha^{\star}\beta & 1\\
2 \alpha^{\star}\beta & 1 & 0 & 2 \alpha^{\star}\beta \\
2 \alpha\beta^{\star} & 0 & 1 & 2 \alpha\beta^{\star} \\
1 & 2 \alpha\beta^{\star} & 2 \alpha^{\star}\beta & 4 |\alpha|^2
\end{array}
\right).
\label{3.5}
\end{eqnarray}
From the fact that one of  the four eigenvalues
\begin{eqnarray}
\left\{ \frac{1}{6}, \frac{1}{6}, \frac{2-\sqrt{5}}{6},
\frac{2+\sqrt{5}}{6}\right\},
\label{3.6}
\end{eqnarray}
of this partially transposed operator is negative for all
values of $\alpha$ (i.e., for arbitrary state of the original qubit)
it follows that the two qubits at the output of the quantum
copier are nonclassically entangled. The fact that the
eigenvalues of the transposed density operator are input-state
independent (and combined with the fact that  distance $d_2$
between $\hat{\rho}_{a_2a_3}^{(out)}$ and $\hat{\rho}_{a_2a_3}^{(id)}$
is also input-state independent)
suggests that the degree of entanglement between the
copied qubits is also input-state independent.

\section{QUANTUM TRIPLICATOR}
When it is {\it a priori} known that the original qubit is initially
in a superposition state (\ref{1.1}) with the mean value of the
observable $\hat{\sigma}_y$ equal to zero (i.e., $\alpha$ and $\beta$
are real) then the quantum copying network presented in Fig.1 can serve
also as a quantum triplicator. That is, out of a single original
qubit this device can create three identical qubits with equal
density operator $\hat{\rho}_{a_j}^{(out)}$, i.e.,
\begin{eqnarray}
\hat{\rho}_{a_1}^{(out)} =
\hat{\rho}_{a_2}^{(out)} =
\hat{\rho}_{a_3}^{(out)},
\label{4.1}
\end{eqnarray}
such that the distances
$d_1(\hat{\rho}_{a_j}^{(out)};\hat{\rho}_{a_j}^{(id)})$
given by (\ref{1.4}) are constant (i.e., they do not depend on
$\alpha$).  This quantum triplicator is input-state independent,
but we have to remember that the class of original qubits for
which this is true is restricted.

The triplicator network is exactly the same as the one considered
in the previous section except we have to perform the rotation
$\hat{R}_3(\theta_2)$ in the opposite direction. That is the angles
$\theta_1$ and $\theta_2$ are the same as specified by Eq.(\ref{2.9}), but
$\theta_2= {\rm arcsin}\left(1/2-\sqrt{2}/3\right)^{1/2}$. In this case
the state  $|\Psi\rangle_{a_2a_3}^{(prep)}$ reads
\begin{eqnarray}
|\Psi\rangle_{a_2a_3}^{(prep)}= \frac{1}{\sqrt{12}}
\left(3|00\rangle +|01\rangle +|10\rangle
+|11\rangle\right).
\label{4.2}
\end{eqnarray}
With the help of Eq.(\ref{2.10}) we now find the output state of
the quantum triplicator:
\begin{eqnarray}
|\Psi\rangle_{a_1a_2a_3}^{(out)}= \frac{1}{\sqrt{12}}
\left(3\alpha|000\rangle +\alpha|101\rangle +\alpha|110\rangle
+\alpha|011\rangle + 3\beta|111\rangle	+\beta|010\rangle
+\beta|001\rangle +\beta|100\rangle
\right).
\label{4.3}
\end{eqnarray}
When $\alpha$ and $\beta$ are real then we find that the three qubits
at the output of the triplicator have {\it identical} density operators
given by Eq.(\ref{1.11}) with the scaling factor $s=2/3$.

Moreover, we find that the three two-qubit density operators
at the output of the triplicator are mutually equal. In the matrix form
they read
\begin{eqnarray}
\hat{\rho}_{a_2a_3}^{(out)}=
\hat{\rho}_{a_1a_2}^{(out)}=
\hat{\rho}_{a_1a_3}^{(out)}=
\frac{1}{12}
\left(
\begin{array}{cccc}
8 \beta^2 +1 & 4 \alpha\beta & 4 \alpha\beta & 3\\
4 \alpha\beta & 1 & 1 & 4 \alpha\beta \\
4 \alpha\beta & 1 & 1 & 4 \alpha\beta \\
3 & 4 \alpha\beta & 4 \alpha\beta & 8 \alpha^2+1
\end{array}
\right).
\label{4.4}
\end{eqnarray}
This quantum triplicator  operates in such way that all distances
between output qubits and ideal copies, i.e.
\begin{eqnarray}
d_1(\hat{\rho}_{a_j}^{(out)};\hat{\rho}_{a_j}^{(id)})=\frac{1}{18};
\qquad
d_2(\hat{\rho}_{a_ja_l}^{(out)};\hat{\rho}_{a_ja_l}^{(id)})=\frac{2}{9};
\qquad
d_3(\hat{\rho}_{a_1a_2a_3}^{(out)};\hat{\rho}_{a_1a_2a_3}^{(id)})=\frac{1}{2},
\label{4.5}
\end{eqnarray}
are constant for arbitrary real values of $\alpha$.

We note that the two-qubit density matrices (\ref{4.4}) are inseparable,
because one of the eigenvalues
\begin{eqnarray}
\left\{-\frac{1}{6},\frac{1}{3},\frac{5+\sqrt{17}}{12},
\frac{5-\sqrt{17}}{12}
\right\}
\label{4.6}
\end{eqnarray}
of the corresponding partially transposed matrix is negative. Because
this negative eigenvalue does not depend on $\alpha$, and the fact that
$d_2$ is input-state independent, we can conclude that quantum
triplicator creates a specific class of two-qubit states characterized by the
same degree of entanglement.

Next we turn our attention to the fact that the scaling factor
$s=2/3$, which relates the output qubits to the original qubit,
is in our case larger than that found in Gisin's triplication procedure
\cite{Gisin},
where it is $s=5/9$. While our scaling factor is larger, there is a price
to pay.
Namely, our triplication network requires {\it a priori} knowledge;
the original qubit must be described by the state
vector (\ref{1.1}) with real $\alpha$ and $\beta$. Gisin's scheme is more
general, because it triplicates all qubits (\ref{1.1}) and the
quality of the copies is independent of the input state.   However, the
quality
of his copies is not as good which can be seen directly
from the fact that for his procedure
the distance between the copied and original qubits
$d_1(\hat{\rho}_{a_j}^{(out)};\hat{\rho}_{a_j}^{(id)})$ is almost two-times
(to be precise, 16/9-times) larger than with ours.
In fact, there exist a general tradeoff between the {\it a priori}
knowledge of the state of the original qubit and the quality
of the copying: the better we know the initial state of the original qubit
the better copying transformation can be. For example, if
we know exactly the state of the original qubit, we can produce as
many perfect copies as we want.

Finally we analyze the output state of the triplicator
network described in Fig.1 when the original qubit is in
an arbitrary superposition state (\ref{1.1}) (with $\alpha$ and
$\beta$ complex). Using the general expression (\ref{4.3})
for the output of the triplicator we find that the individual
qubits at the output are equal, i.e.
$ \hat{\rho}_{a_1}^{(out)} = \hat{\rho}_{a_2}^{(out)} =
\hat{\rho}_{a_3}^{(out)}$, with the density matrices given by the
expression
\begin{eqnarray}
\hat{\rho}_{a_j}^{(out)}=
\frac{1}{6}
\left(
\begin{array}{cc}
4| \beta|^2 +1 & 3 \alpha^{\star}\beta + \alpha\beta^{\star}\\
 3 \alpha\beta^{\star} + \alpha^{\star}\beta
 & 4 |\alpha|^2+1
\end{array}
\right); \qquad j=1,2,3.
\label{4.7}
\end{eqnarray}
In general, these density operators cannot be written in the scaled
form (\ref{1.11}) and consequently, the distance between the
output and input qubits depends on the initial state of the
original qubit, i.e.
\begin{eqnarray}
d_1(\hat{\rho}_{a_j}^{(out)};\hat{\rho}_{a_j}^{(id)})=\frac{1}{18}
\left(1+ 12 |\alpha|^2 |\beta|^2 \sin^2 \varphi\right).
\label{4.8}
\end{eqnarray}

The two-qubit density operators at the output of the triplicator
are also equal, and they can be described by the
density matrix
\begin{eqnarray}
\hat{\rho}_{a_2a_3}^{(out)} = \hat{\rho}_{a_1a_2}^{(out)}  =
\hat{\rho}_{a_1a_3}^{(out)}  =\frac{1}{12}
\left(
\begin{array}{cccc}
8|\beta|^2 +1 & 3 \alpha^{\star}\beta +\alpha\beta^{\star} &
3 \alpha^{\star}\beta +\alpha\beta^{\star} & 3\\
3 \alpha\beta^{\star} +\alpha^{\star}\beta & 1 & 1 &
3 \alpha^{\star}\beta +\alpha\beta^{\star} \\
3 \alpha\beta^{\star} +\alpha^{\star}\beta  & 1 & 1 &
3 \alpha^{\star}\beta +\alpha\beta^{\star} \\
3 & 3 \alpha\beta^{\star} +\alpha^{\star}\beta &
3 \alpha\beta^{\star} +\alpha^{\star}\beta & 8|\alpha|^2 +1
\end{array}
\right).
\label{4.9}
\end{eqnarray}
From this expression we can easily find that the
 two-qubit distances
$d_2(\hat{\rho}_{a_ka_l}^{(out)};\hat{\rho}_{a_ka_l}^{(id)})$ between the
actual output of the triplicator and the ideal case  are input-state
dependent, i.e.
\begin{eqnarray}
d_2(\hat{\rho}_{a_ka_l}^{(out)};\hat{\rho}_{a_ka_l}^{(id)}) =\frac{2}{9}
\left(1 + 12|\alpha|^2 |\beta|^2 \sin^2\varphi\right);
\qquad k,l=1,2,3;\qquad k\neq l.
\label{4.10}
\end{eqnarray}
Analogously for the three-qubit distance
$d_3(\hat{\rho}_{a_1a_2a_3}^{(out)};\hat{\rho}_{a_1a_2a_3}^{(id)})$
we find
\begin{eqnarray}
d_3(\hat{\rho}_{a_1a_2a_3}^{(out)};\hat{\rho}_{a_1a_2a_3}^{(id)})=
\frac{1}{2}
\left(1 +12 |\alpha|^2 |\beta|^2 \sin^2\varphi\right).
\label{4.11}
\end{eqnarray}
Here the minimum values of the distances $d_j$ ($j=1,2,3$) are obtained
when $\varphi=0,\pi$ and in this case they do not depend
on the particular value of $|\alpha|$.

From the explicit expression (\ref{4.9}) we find that
the two-qubit density matrices
are inseparable
for an arbitrary state of the input qubit. This means
that quantum triplication ``creates'' very specific quantum correlations
between the output qubits. Namely, one of the eigenvalues of the
partially transposed matrix (\ref{4.9}) is negative for {\em arbitrary}
values
$|\alpha|$ and $\varphi$. Moreover, there exists correspondence
between the behavior of the distance
$d_1(\hat{\rho}_{a_j}^{(out)};\hat{\rho}_{a_j}^{(id)})$
and the value of the negative eigenvalue $E$ of the partially
transposed matrix. In particular, when $\varphi=0,\pi$
then this  eigenvalue does not depend on $|\alpha|$ and is equal to $-1/6$.
The corresponding distance $d_1$ in this case is minimal and equal
to $1/18$ (irrespective of $|\alpha|$).
As the distance  $d_1$ increase this eigenvalue  decreases.
Specifically, for a given value of $|\alpha|$ the distance $d_1$
is maximal when $\varphi=\pi/2$. Correspondingly, the negative eigenvalue
$E$ of the partially transposed matrix for a given $|\alpha|$ takes its
minimal value when $\varphi=\pi/2$. In this case $E$ can be approximated
by its upper bound $\bar{E}$
\begin{eqnarray}
E\lesssim \bar{E} = - \left(\frac{1 + 4 (\sqrt{5}-2) |\alpha|^2 |\beta|^2}{6}\right),
\label{4.12}
\end{eqnarray}
which clearly reveals a dependence between the distance $d_1$
[given by Eq.(\ref{4.8}) with $\varphi=\pi/2$]	and
the negative eigenvalue $E$.
This observation suggests
that the copying schemes analogous to the triplication network
discussed above can serve  as specific ``quantum entanglers'' and that
the measure of entanglement can be operationally related to
a specific distance $d_1$.

\section{CONCLUSION}
It is possible to construct devices which copy the information
in a quantum state as long as one does not demand perfect
copies.  One can build either a duplicator, which
produces two copies, or a triplicator, which produces three.  Both
of these devices can be realized by simple networks of quantum
gates, which should make it possible to construct them in the
laboratory.

There are a number of unanswered questions about quantum copiers.
Perhaps the most obvious is which quantum copier is the best.
Recently it has been shown \cite{Bruss}
that the UQCM described in this paper
is the best quantum copier able to produce two copies of the original
qubit. It is not known, however, how to construct
the best quantum triplicator (or, in general, a device which will
produce multiple copies, the so-called multiplicator).
 There exist bounds on how well one can
do, which follow from unitarity, but they are not realized by
existing copiers \cite{Hillery}.  This is at least partially the fault of the
bounds which are probably lower than they have to be.

A quantum copier takes quantum information in one system and spreads
it among several.  It would be nice to be able to see how this happens
qualitatively, but, at the moment, it is not clear how to do this.  The
problem is that we are interested in how only a part of the information
flows through the machine.  It is only the information in the input
state, and not that in the two input qubits, which enter the machine in
standard states, the so-called ``blank pieces of paper'', which matters,
but it seems to be difficult to separate the effect of the two in
the action of the machine.

This issue is connected to another, which is how to best use the copies
to gain information about the input state.  In a previous paper we
showed how nonselective measurements of a single quantity on one
of the copies can be used
to gain information about the original and leave the one-particle
reduced density matrix of the other copy unchanged.  An interesting
extension of this would be to ask, for a given number of copies, how
much information we can gain about the original state by performing
different kinds of measurements on the copies.

It is clear that quantum copying still presents both theoretical
and experimental challenges.  We hope to be able to address some
of issues raised by the questions in the preceding paragraphs
in future publications.

\vspace{1.5truecm}

{\bf Acknowledgements}\newline
We thank Nicolas Gisin for communication of
his resent results to us. We also thank Peter Knight and Artur Ekert
for useful discussions.
This work was supported  by the United Kingdom Engineering
and Physical Sciences Research Council, by the grant agency VEGA
of the Slovak Academy of Sciences (under the project
2/1152/96), and  by the National Science
Foundation under the grant  INT 9221716.

\section*{FIGURE CAPTION}
{\bf FIG. 1} Graphical representation of the UQCM network.  The logical
controlled-NOT $\hat{P}_{kl}$ given by Eq.(\ref{2.2})
has as its input a control qubit
(denoted as $\bullet$ ) and  a target qubit  (denoted as $\circ$ ).
The action of the single-qubit operator {\bf R} is specified by the
transformation (\ref{2.1}).
We separate the preparation of the quantum copier from the
copying process itself. The copying, i.e. the transfer of quantum
information from the original qubit, is performed by a sequence of
four controlled-NOTs. We note that the amplitude information from the
original qubit is copied in the obvious direction in an XOR or
the controlled-NOT
operation. Simultaneously, the phase information is copied in the
opposite direction making the XOR a simple model of quantum non-demolition
measurement and its back-action.

\newpage

\newcounter{cms}
\setlength{\unitlength}{1.0mm}

\vskip3.0truecm

\begin{center}
\begin{picture}(118,105)
\put(-15,70){\makebox(0,0)[c]{$|\Psi\rangle_{a_1}^{(in)}$}}
\put(4,60){\makebox(0,0)[c]{$|0\rangle_{a_2}$}}
\put(4,50){\makebox(0,0)[c]{$|0\rangle_{a_3}$}}
\put(-7,70){\line(1,0){88}}
\put(83,70){\line(1,0){11}}
\put(96,70){\line(1,0){18}}
\put(8,60){\line(1,0){8}}
\put(20,60){\line(1,0){4}}
\put(26,60){\line(1,0){12}}
\put(40,60){\line(1,0){4}}
\put(48,60){\line(1,0){8}}
\put(58,60){\line(1,0){23}}
\put(83,60){\line(1,0){31}}
\put(8,50){\line(1,0){16}}
\put(26,50){\line(1,0){4}}
\put(34,50){\line(1,0){4}}
\put(40,50){\line(1,0){28}}
\put(70,50){\line(1,0){24}}
\put(96,50){\line(1,0){18}}
\put(82,70){\circle{2}}
\put(95,70){\circle{2}}
\put(18,60){\circle{4}}
\put(39,60){\circle{2}}
\put(46,60){\circle{4}}
\put(18,60){\makebox(0,0)[c]{\bf R}}
\put(46,60){\makebox(0,0)[c]{\bf R}}
\put(57,60){\circle{2}}
\put(69,50){\circle{2}}
\put(25,50){\circle{2}}
\put(32,50){\circle{4}}
\put(39,50){\circle{2}}
\put(32,50){\makebox(0,0)[c]{\bf R}}
\put(57,70){\circle*{2}}
\put(69,70){\circle*{2}}
\put(25,60){\circle*{2}}
\put(82,60){\circle*{2}}
\put(39,50){\circle*{2}}
\put(95,50){\circle*{2}}
\put(25,51){\line(0,1){8}}
\put(39,51){\line(0,1){8}}
\put(57,61){\line(0,1){8}}
\put(82,61){\line(0,1){8}}
\put(69,51){\line(0,1){18}}
\put(95,51){\line(0,1){18}}
%
%\put(67.5,20){\vector(0,1){10}}
%\put(30,30){\makebox(0,0)[c]{ preparation \hskip2.0truecm  copying}}
\put(13,45){\line(1,0){37}}
\put(13,45){\line(0,1){20}}
\put(50,45){\line(0,1){20}}
\put(13,65){\line(1,0){37}}
\put(52,45){\line(1,0){49}}
\put(52,45){\line(0,1){30}}
\put(101,45){\line(0,1){30}}
\put(52,75){\line(1,0){49}}
\put(-2,35){\line(1,0){110}}
\put(-2,35){\line(0,1){45}}
\put(108,35){\line(0,1){45}}
\put(-2,80){\line(1,0){110}}
\put(22,41){\mbox{preparation}}
\put(69,41){\mbox{copying}}
\put(40,29){\mbox{quantum copier}}
%\Huge
%\put(83,65){\makebox(0,0)[c]{$\}$}}
%\normalsize
\put(122,60){\makebox(0,0)[c]{$|\Psi\rangle_{a_1a_2a_3}^{(out)}$}}
%\put(131,55){\makebox(0,0)[c]{$\hat{\rho}_{a_1a_2}^{(out)}$}}

\end{picture}\\

\end{center}

\vskip10.0truecm

\begin{center}
FIGURE 1.
\end{center}

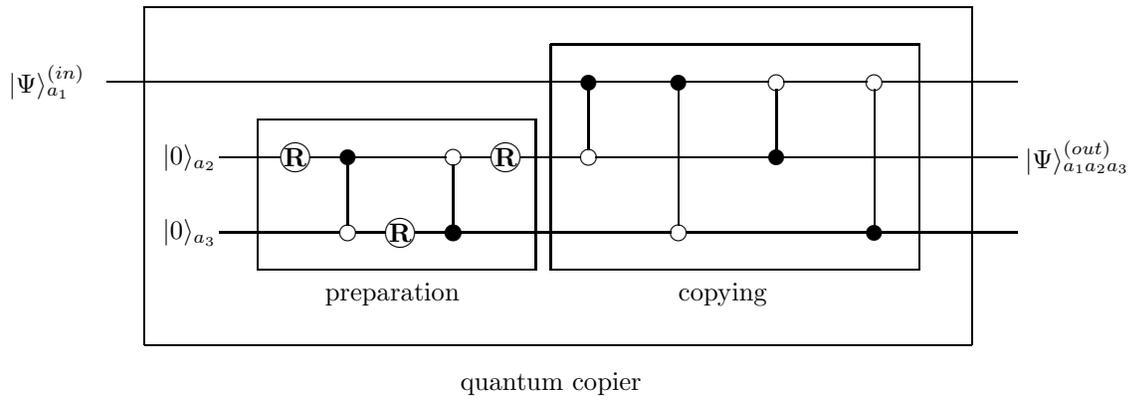
\end{document}